\documentclass{PoS}
\usepackage{amsmath}

\def\slashb#1{\not\!\!#1}

\title{A direct relation between confinement and chiral symmetry breaking 
in temporally odd-number lattice QCD}

\ShortTitle{Relation between confinement \& chiral symmetry breaking 
in temporally odd lattice QCD}

\author{\speaker{Takahiro M. Doi}, Hideo Suganuma \\
        Department of Physics \& Division of Physics and Astronomy, 
Graduate School of Science, \\
Kyoto University, 
Kitashirakawaoiwake, Sakyo, Kyoto 606-8502, Japan\\
        E-mail: \email{doi@ruby.scphys.kyoto-u.ac.jp}}

\author{Takumi Iritani \\
High Energy Accelerator Research Organization (KEK), 
Tsukuba, Ibaraki 305-0801, Japan}

\abstract{
In the lattice QCD formalism, we derive a gauge-invariant analytical relation 
connecting the Polyakov loop and the Dirac modes on a temporally odd-number 
lattice, where the temporal lattice size is odd, 
with the normal (nontwisted) periodic boundary condition. 
This analytical relation indicates that low-lying Dirac modes have 
little contribution to the Polyakov loop. 
Using lattice QCD simulations, 
we numerically confirm the analytical relation and 
the negligible contribution of low-lying Dirac modes 
to the Polyakov loop at the quenched level, i.e., 
the Polyakov loop is almost unchanged by removing low-lying Dirac-mode contribution 
from the QCD vacuum generated by lattice QCD in both confinement and deconfinement phases. 
Thus, we conclude that there is no one-to-one correspondence 
between confinement and chiral symmetry breaking in QCD. 
As a new method, modifying the Kogut-Susskind formalism, 
we develop a method for spin-diagonalizing the Dirac operator 
on the temporally odd-number lattice. 
}

\FullConference{31st International Symposium on Lattice Field Theory - LATTICE 2013\\
		July 29 - August 3, 2013\\
		Mainz, Germany}

\begin{document}

\section{Introduction}
Color confinement and chiral symmetry breaking have been investigated 
as interesting non-perturbative phenomena in low-energy QCD 
in many analytical and numerical studies. 
However, their properties are not sufficiently understood directly from QCD. 
The Polyakov loop is an order parameter for quark confinement \cite{Rothe}. 
At the quenched level, the Polyakov loop is the exact order parameter for 
quark confinement, and its expectation value is zero in confinement phase 
and is nonzero in deconfinement phase. 
As for the chiral symmetry, 
low-lying Dirac modes are essential for chiral symmetry breaking in QCD, 
according to the Banks-Casher relation \cite{BC}.

Not only the properties of confinement and chiral symmetry breaking in QCD 
but also their relation is an interesting challenging subject \cite{Suganuma}.
By removing QCD monopoles in the maximally Abelian gauge, 
both confinement and chiral symmetry breaking are lost 
in lattice QCD \cite{Miyamura}.
The transition temperatures of deconfinement and chiral restoration 
are almost same in finite temperature QCD \cite{Karsch}. 
From these facts, it is suggested that 
confinement and chiral symmetry breaking are strongly correlated.
In recent lattice-QCD numerical studies, however, 
it is found that confinement properties 
do not change by removing low-lying Dirac modes from the QCD vacuum, 
which means no one-to-one correspondence between 
confinement and chiral symmetry breaking in QCD \cite{GIS}.

In this study, we derive an analytical relation connecting 
the Polyakov loop and the Dirac modes on a temporally odd-number lattice, 
%using the gauge-invariant Dirac-mode expansion, 
and discuss the relation between confinement and chiral symmetry breaking.
As a by-product, 
we develop a new method for spin-diagonalizing the Dirac operator 
on the temporally odd-number lattice by modifying the Kogut-Susskind (KS) 
formalism. 
Using this method, we numerically confirm the analytical relation.

\section{Dirac modes in lattice QCD}
In this section, we review the Dirac operator, 
its eigenvalues and its eigenmodes (Dirac modes) in ${\rm SU}(N_{\rm c})$ 
lattice QCD \cite{GIS}.
We use a standard square lattice with spacing $a$, and the notation of 
sites $s=(s_1, s_2, s_3, s_4) \ (s_\mu=1,2,\cdots,N_\mu)$, 
and link-variables $U_\mu(s)={\rm e}^{iagA_\mu(s)}$ 
with gauge fields $A_\mu(s) \in su(N_c)$ and gauge coupling $g$.
In lattice QCD, the Dirac operator $\slashb{D}=\gamma_\mu D_\mu$ 
is given by
\begin{eqnarray}
 \slashb{D}_{s,s'} 
      = \frac{1}{2a} \sum_{\mu=1}^4 \gamma_\mu 
\left[ U_\mu(s) \delta_{s+\hat{\mu},s'}
        - U_{-\mu}(s) \delta_{s-\hat{\mu},s'} \right], \label{Eq:DiracOp}
\end{eqnarray}
with $U_{-\mu}(s)\equiv U^\dagger_\mu(s-\hat\mu)$.
Here, $\hat\mu$ is the unit vector in direction $\mu$ in the lattice unit.
In this paper, we define all the $\gamma$-matrices to be hermite as 
$\gamma_\mu^\dagger=\gamma_\mu$.
Since the Dirac operator is anti-hermite 
in this definition of $\gamma_\mu$, 
the Dirac eigenvalue equation is expressed as
\begin{eqnarray}
\slashb{D}|n\rangle =i\lambda_n|n \rangle
\end{eqnarray}
with the Dirac eigenvalue $i\lambda_n$ ($\lambda_n \in {\bf R}$) 
and the Dirac eigenstate $|n \rangle$.
Note that the chiral partner $\gamma_5|n\rangle$ is also 
an eigenstate with the eigenvalue $-i\lambda_n$.
Using the Dirac eigenfunction $\psi_n(s)\equiv\langle s|n \rangle $, 
the explicit form for the Dirac eigenvalue equation is written by 
\begin{eqnarray}
 \frac{1}{2a}& \sum_{\mu=1}^4 \gamma_\mu
[U_\mu(s)\psi_n(s+\hat \mu)-U_{-\mu}(s)\psi_n(s-\hat \mu)]=i\lambda_n \psi_n(s). \label{Eq:eigenEqExplicit}
\end{eqnarray}

\section{Operator formalism in lattice QCD}
In this section, we present operator formalism in lattice QCD \cite{GIS}.
First, we define the link-variable operator $\hat{U}_{\pm\mu}$ 
by the matrix element,
\begin{eqnarray}
\langle 
s | \hat{U}_{\pm\mu} |s' \rangle=U_{\pm\mu}(s)\delta_{s\pm\hat{\mu},s'}.
\end{eqnarray}
Using the link-variable operator, 
the Polyakov loop $\langle L_P \rangle$ is expressed as 
\begin{eqnarray}
 \langle L_P \rangle=\frac{1}{3V}\langle {\rm Tr}_c \{\hat U_4^{N_4}\} \rangle
=\frac{1}{3V}\langle \sum_s {\rm tr}_c
\{U_4(s)U_4(s+\hat t)U_4(s+2\hat t)\cdots U_4(s+(N_4-1)\hat t)\}
\rangle, \label{Eq:LOp}
\end{eqnarray}
with the 4D lattice volume $V=N_1N_2N_3N_4$. 
Here, ``Tr$_c$'' denotes the functional trace of 
${\rm Tr}_c \equiv \sum_s {\rm tr}_c$
with the trace ${\rm tr}_c$ over color index. 
In this formalism, the Dirac operator is simply expressed as
\begin{eqnarray}
 \hat{\slashb{D}}=
\frac{1}{2a}\sum_{\mu=1}^4
\gamma_\mu(\hat{U}_\mu-\hat{U}_{-\mu}). \label{Eq:DOp}
\end{eqnarray}

\vspace{-0.45cm}

\section{A direct analytical relation between the Polyakov loop 
and Dirac modes in temporally odd-number lattice QCD}

We consider a temporally odd-number lattice, 
where the temporal lattice size $N_4$ is odd, 
with the normal (nontwisted) periodic boundary condition 
in both temporal and spatial directions. 
The spatial lattice size $N_{1 \sim 3} (> N_4)$ is taken to be even.
As a key quantity, we first introduce 
\begin{eqnarray}
I\equiv {\rm Tr}_{c,\gamma} (\hat{U}_4\hat{\slashb{D}}^{N_4-1}) \label{Eq:I}
\end{eqnarray}
with the functional trace 
${\rm Tr}_{c,\gamma}\equiv \sum_s {\rm tr}_c 
{\rm tr}_\gamma$ including also 
the trace ${\rm tr}_\gamma$ over spinor index.
Its expectation value 
\begin{eqnarray}
 \langle I\rangle=\langle {\rm Tr}_{c,\gamma} (\hat{U}_4\hat{\slashb{D}}^{N_4-1})\rangle 
\label{Eq:VEVI}
\end{eqnarray}
is obtained as the gauge-configuration average in lattice QCD.
In the case of large volume $V$, one can expect 
$\langle O \rangle \simeq {\rm Tr} O/{\rm Tr} 1$ 
for any operator at each gauge configuration.

From Eq.(\ref{Eq:DOp}), $\hat{U}_4\hat{\slashb{D}}^{N_4-1}$ is expressed as 
a sum of products of $N_4$ link-variable operators. Since $N_4$ is odd, 
$\hat{U}_4\hat{\slashb{D}}^{N_4-1}$ does not have any closed loops 
except for the term proportional to $\hat{U}_4^{N_4}$. 
Therefore, according to Elitzur's theorem and using Eq.(\ref{Eq:LOp}), 
we obtain 
\begin{eqnarray}
\langle I\rangle=\frac{1}{(2a)^{N_4-1}}
\langle {\rm Tr}_{c,\gamma} \{\hat{U}_4^{N_4}\}\rangle
=\frac{4}{(2a)^{N_4-1}}
\langle {\rm Tr}_{c} \{\hat{U}_4^{N_4}\}\rangle
=\frac{12V}{(2a)^{N_4-1}}\langle L_P \rangle. \label{Eq:I1}
\end{eqnarray}
On the other hand, by performing the functional trace 
in Eq.(\ref{Eq:VEVI}) with the 
Dirac mode basis $|n\rangle$ satisfying $\sum_n |n\rangle \langle n|=1$, 
we find
\begin{eqnarray}
 \langle I\rangle=\sum_n\langle n|\hat{U}_4\slashb{\hat{D}}^{N_4-1}|n\rangle
=i^{N_4-1}\sum_n\lambda_n^{N_4-1}\langle n|\hat{U}_4| n \rangle.  \label{Eq:I2}
\end{eqnarray}
Combing Eqs. (\ref{Eq:I1}) and (\ref{Eq:I2}), we obtain the relation between 
the Polyakov loop $ \langle L_P \rangle$ 
and the Dirac eigenvalues $i\lambda_n$: 
\begin{eqnarray}
 \langle L_P \rangle=\frac{(2ai)^{N_4-1}}{12V}
\sum_n\lambda_n^{N_4-1}\langle n|\hat{U}_4| n \rangle.  \label{Eq:RelOrig}
\end{eqnarray}
This is a relation directly connecting the Polyakov loop and the Dirac modes, 
i.e., a Dirac spectral representation of the Polyakov loop, 
and is valid on the temporally odd-number lattice. 
From this relation  (\ref{Eq:RelOrig}), 
we can investigate each Dirac-mode contribution 
to the Polyakov loop individually.

From Eq.(\ref{Eq:RelOrig}), we can discuss the relation between confinement 
and chiral symmetry breaking in QCD. 
Because of the factor $\lambda_n^{N_4-1}$, the contribution from 
low-lying Dirac-modes with $|\lambda_n|\simeq 0$ 
is very small in the sum of RHS in Eq.(\ref{Eq:RelOrig}),
compared to the other Dirac-mode contribution. 
In fact, the low-lying Dirac modes have little contribution 
to the Polyakov loop. This is consistent with the previous numerical 
lattice result that confinement properties are almost unchanged 
by removing low-lying Dirac modes from the QCD vacuum \cite{GIS}. 
Thus, we conclude from the relation (\ref{Eq:RelOrig}) that there is 
no one-to-one correspondence between confinement and chiral symmetry breaking.

\section{Modified KS formalism for temporally odd-number lattice}
The Dirac operator $\slashb{D}$ has a large dimension of 
$(4 \times N_{\rm c}\times V)^2$, so that 
the numerical cost for solving the Dirac eigenvalue equation is quite huge.
This numerical cost can be partially reduced 
using the Kogut-Susskind (KS) formalism \cite{Rothe,GIS,KS}.
However, the original KS formalism can be applied only to the ``even lattice'' 
where all the lattice sizes $N_\mu$ are even number. 
In this section, we modify the KS formalism 
to be applicable to the odd-number lattice. 
Using the modified KS formalism, we can reduce the numerical cost 
in the case of the temporally odd-number lattice.

First, we recall the original KS formalism for even lattices.
Using the matrix 
\begin{eqnarray}
 T(s)\equiv\gamma_1^{s_1}\gamma_2^{s_2}\gamma_3^{s_3}\gamma_4^{s_4}, 
\label{Eq:T}
\end{eqnarray}
all the $\gamma$-matrices can be diagonalized as 
\begin{eqnarray}
 T^\dagger(s)\gamma_\mu T(s\pm\hat{\mu})=\eta_\mu(s){\bf 1}, \label{Eq:TgammaT}
\end{eqnarray}
where $\eta_\mu(s)$ is the staggered phase,
\begin{eqnarray}
 \eta_1(s)\equiv 1, \ \ \eta_\mu(s)\equiv (-1)^{s_1+\cdots+s_{\mu-1}} 
\ (\mu \geq 2). \label{Eq:eta}
\end{eqnarray}
Then, one can spin-diagonalize the Dirac operator $\slashb{D}$ as 
\begin{eqnarray}
 \sum_\mu T^\dagger(s) \gamma_\mu D_\mu T(s+ \hat \mu) = {\rm diag}(\eta_\mu D_\mu,\eta_\mu D_\mu,\eta_\mu D_\mu,\eta_\mu D_\mu), \label{Eq:TDiracT}
\end{eqnarray}
where $\eta_\mu D_\mu$ is the KS Dirac operator given by
\begin{eqnarray}
(\eta_\mu D_\mu)_{ss'}=\frac{1}{2a}\sum_{\mu=1}^{4}\eta_\mu(s)\left[U_\mu(s)\delta_{s+\hat{\mu},s'}-U_{-\mu}(s)\delta_{s-\hat{\mu},s'}\right]. \label{Eq:KSDiracOp}
\end{eqnarray}
Equation (\ref{Eq:TDiracT}) shows fourfold degeneracy of the Dirac eigenvalue 
relating to the spiror structure, and 
then all the eigenvalues $i\lambda_n$ are obtained by solving 
the reduced Dirac eigenvalue equation
\begin{eqnarray}
\eta_\mu D_\mu|n) =i\lambda_n|n ). \label{Eq:KSEigenEqOp}
\end{eqnarray}
Using the eigenfunction $\chi_n(s)\equiv\langle s|n )$ 
of the KS Dirac operator, 
the explicit form of Eq.(\ref{Eq:KSEigenEqOp}) reads
\begin{eqnarray}
\frac{1}{2a}\sum_{\mu=1}^4 
\eta_\mu(x)[U_\mu(x) \chi_n(x+\hat \mu)-U_{-\mu}(x)
\chi_n(x-\hat \mu)] =i\lambda_n\chi_n(x), \label{Eq:KSEigenEq}
\end{eqnarray}
where the relation between the Dirac eigenfunction 
$\psi_n(s)$ and the spinless eigenfunction $\chi_n(s)$ is
\begin{eqnarray}
\psi_n(s)=T(s)\chi_n(s). \label{Eq:PsiChiEven}
\end{eqnarray}
%In fact, there are four Dirac eigenfunctions $\psi_n(s)$ for each 
%eigenvalue $\lambda_n$ because of fourfold degeneracy of the Dirac eigenvalues.

Note here that the original KS formalism is applicable only to even lattices 
in the presence of the periodic boundary condition. 
In fact, the periodic boundary condition requires 
\begin{eqnarray}
T(s+N_\mu\hat{\mu})=T(s) \ (\mu=1,2,3,4),
\end{eqnarray}
however, it is satisfied only on even lattices.
Note also that, while the spatial boundary condition can be changed arbitrary, 
the temporal periodic boundary condition physically appears and 
cannot be changed at finite temperatures. 
Thus, the original KS formalism cannot be applied 
on the temporally odd-number lattice.

Now, we consider the temporally odd-number lattice, 
with all the spatial lattice size being even. 
Instead of the matrix $T(s)$, we introduce a new matrix 
\begin{eqnarray}
 M(s)\equiv\gamma_1^{s_1}\gamma_2^{s_2}\gamma_3^{s_3}\gamma_4^{s_1+s_2+s_3}, 
\label{Eq:M}
\end{eqnarray}
where the exponent of $\gamma_4$ differs from $T(s)$ in Eq.(\ref{Eq:T}).
As a remarkable feature, the requirement from 
the periodic boundary condition is satisfied 
on the temporally odd-number lattice:
\begin{eqnarray}
M(s+N_\mu\hat{\mu})=M(s) \ (\mu=1,2,3,4).
\end{eqnarray}
Using the matrix $M(s)$, all the $\gamma$-matrices 
are transformed to be proportional to $\gamma_4$:
\begin{eqnarray}
 M^\dagger(s)\gamma_\mu M(s\pm\hat{\mu})=\eta_\mu(s)\gamma_4, 
\label{Eq:MgammaM}
\end{eqnarray}
where $\eta_\mu(x)$ is the staggered phase given by Eq.(\ref{Eq:eta}).
In the Dirac representation, $\gamma_4$ is diagonal as 
\begin{eqnarray}
 \gamma_4={\rm diag}(1,1,-1,-1) \ \ \ (\rm{Dirac \ representation}),
\label{Eq:gamma_4}
\end{eqnarray}
and we take the Dirac representation.
Thus, we can spin-diagonalize the Dirac operator $\slashb{D}$ in the case of 
the temporally odd-number lattice:
\begin{eqnarray}
 \sum_\mu M^\dagger(s) \gamma_\mu D_\mu M(s+ \hat \mu) = {\rm diag}(\eta_\mu D_\mu,\eta_\mu D_\mu,-\eta_\mu D_\mu,-\eta_\mu D_\mu), \label{Eq:MDiracM}
\end{eqnarray}
where $\eta_\mu D_\mu$ is 
the KS Dirac operator given by Eq.(\ref{Eq:KSDiracOp}).
Then, for each $\lambda_n$, 
two positive modes and two negative modes appear 
relating to the spinor structure on the temporally odd-number lattice.
(Note also that the chiral partner $\gamma_5 |n \rangle$ 
gives an eigenmode with the eigenvalue $-i\lambda_n$.)
%
%Dirac eigenvalues distribute symmetrically.
%
In any case, all the eigenvalues $i\lambda_n$ can be obtained by 
solving the reduced Dirac eigenvalue equation 
\begin{eqnarray}
\eta_\mu D_\mu|n) =\pm i\lambda_n|n ) \label{Eq:KSEigenEqOpOdd}
\end{eqnarray}
just like the case of even lattices.
The relation between the Dirac eigenfunction $\psi_n(s)$ and 
the sponless eigenfunction $\chi_n(s)\equiv\langle s|n ) $ 
is given by 
\begin{eqnarray}
\psi_n(s)=M(s)\chi_n(s) \label{Eq:PsiChiOdd}
\end{eqnarray}
on the temporally odd-number lattice.
%In fact, there are two Dirac eigenfunctions $\psi_n(s)$ for each independent 
%eigenvalues $\lambda_n$ because of twofold degeneracy of Dirac eigenvalues.

\section{Numerical confirmation for the relation between Polyakov loop 
and Dirac modes}

Using the modified KS formalism, Eq.(\ref{Eq:RelOrig}) is rewritten as
\begin{eqnarray}
\langle L_P \rangle =\frac{(2ai)^{N_4-1}}{3V}
\sum_n\lambda_n^{N_4-1}( n|\hat{U}_4| n ). \label{Eq:RelKS}
\end{eqnarray}
Note that the (modified) KS formalism is an exact method for diagonalizing 
the Dirac operator and is not an approximation, so that 
Eqs.(\ref{Eq:RelOrig}) and (\ref{Eq:RelKS}) are completely equivalent.
In fact, the relation (\ref{Eq:RelOrig}) can be confirmed 
by the numerical test of the relation (\ref{Eq:RelKS}). 

We numerically calculate LHS and RHS of the relation (\ref{Eq:RelKS}), 
respectively, and compare them. 
We perform SU(3) lattice QCD Monte Carlo simulations with 
the standard plaquette action at the quenched level 
in both cases of confinement and deconfinement phases.
For the confinement phase, we use $10^3\times 5$ lattice with 
$\beta\equiv2N_{\rm c}/g^2=5.6$ (i.e., $a\simeq 0.25~{\rm fm}$), 
corresponding to $T\equiv1/(N_4a)\simeq160~{\rm MeV}$.
For the deconfinement phase, we use $10^3\times 3$ lattice with $\beta=5.7$ 
(i.e., $a\simeq 0.20~{\rm fm}$), corresponding to 
$T\equiv1/(N_4a)\simeq 330~{\rm MeV}$.
For each phase, 
we use 20 gauge configurations, which are taken every 500 sweeps 
after the thermalization of 5,000 sweeps.

As the numerical result, 
comparing LHS and RHS of the relation (\ref{Eq:RelKS}), 
we find that the relation (\ref{Eq:RelKS}) is almost exact 
for each gauge configuration in both confinement and deconfinement phases.
Therefore, of course, the relation (\ref{Eq:RelKS}) 
is satisfied for the gauge-configuration average. 

%\begin{table}[htb]
%\caption{
%$a$a
%Schematic illustration of the stringy excitation of hadrons. 
%The flux-tube picture of hadrons is idealized as 
%the string picture in the infrared region, 
%which is expected to allow ``stringy excitations'' of hadrons. 
%Since this stringy mode is non-quark-origin excitation, 
%it can be regarded as a gluonic excitation. 
%}
%  \begin{tabular}{|c|c|c|c|} \hline
%configuration No. &1&2&3 \\ \hline
%Re$\langle L\rangle$ & .. & .. & .. \\ 
%Im$\langle L\rangle$ & .. &.. &.. \\\hline 
%Re~$S$ & .. & ... & .. \\ 
%Im~$S$ & .. & ... & .. \\ \hline
%  \end{tabular}
%\end{table}
%---Here-------\\
%Table for the numerical results for the left/right hands of the relation\\
%----------\\

Next, we numerically confirm that the low-lying Dirac modes 
have negligible contribution to the Polyakov loop using Eq.(\ref{Eq:RelKS}).
By checking all the Dirac modes, we 
find that the matrix element $(n|\hat U_4|n)$ is generally nonzero.
In fact, for low-lying Dirac modes, 
the factor $\lambda_n^{N_4-1}$ plays a crucial role 
in RHS of Eq.(\ref{Eq:RelKS}).
Since RHS of Eq.(\ref{Eq:RelKS}) is expressed as a sum of 
the Dirac-mode contribution, 
we calculate the Polyakov loop without low-lying Dirac-mode contribution as 
\begin{eqnarray}
 \langle L_P \rangle_{\rm IR\hbox{-}cut} \equiv \frac{(2i)^{N_4-1}}{3V}
\sum_{|\lambda_n|>\Lambda_{\rm IR}}\lambda_n^{N_4-1}( n|\hat{U}_4| n ) 
\label{Eq:IRcut},
\end{eqnarray}
with the infrared (IR) cut $\Lambda_{\rm IR}$ for the Dirac eigenvalue.
The chiral condensate $\langle \bar qq \rangle$ is given by 
\vspace{-0.1cm}
\begin{eqnarray}
 \langle \bar qq\rangle
&=&-\frac{1}{V}{\rm Tr}_{c,\gamma}\frac{1}{\slashb D+m}
=-\frac{1}{V}\sum_n\frac{1}{i\lambda_n+m}
=-\frac{1}{V}\left(\sum_{\lambda_n>0} \frac{2m}{\lambda_n^2+m^2}
+\frac{\nu}{m}\right), 
\label{Eq:qbarq}
\end{eqnarray}
\vspace{-0.1cm}
where $m$ is the current quark mass 
and $\nu$ the total number of zero modes of $\slashb D$.
The chiral condensate without the contribution from the low-lying Dirac-mode  
below IR cut $\Lambda_{\rm IR}$ is given by 
\vspace{-0.05cm}
\begin{eqnarray}
 \langle \bar qq\rangle_{\Lambda_{\rm IR}} = -\frac{1}{V}\sum_{\lambda_n > \Lambda_{\rm IR}} \frac{2m}{\lambda_n^2+m^2}. \label{Eq:qbarqIR}
\end{eqnarray}
\vspace{-0.1cm}
Here, we take the IR cut of $\Lambda_{\rm IR}\simeq0.4 {\rm GeV}$.
In the confined phase, this IR Dirac-mode cut leads to 
\vspace{-0.05cm}
\begin{eqnarray}
\frac{\langle \bar qq\rangle_{\Lambda_{\rm IR}}}
{\langle \bar qq\rangle} \simeq 0.02
\end{eqnarray}
\vspace{-0.1cm}
and almost chiral-symmetry restoration 
in the case of physical current-quark mass, $m\simeq5 {\rm MeV}$.

We find that 
$\langle L_P \rangle\simeq\langle L_P \rangle_{\rm IR\hbox{-}cut}$  
is numerically satisfied for each gauge configuration 
in both confinement and deconfinement phases. 
Table 1 and 2 show a part of the numerical result 
on $\langle L_P \rangle$ and $\langle L_P \rangle_{\rm IR\hbox{-}cut}$ 
for confinement and deconfinement phases, respectively.
In this way, the Polyakov loop is almost unchanged by removing 
the contribution from the low-lying Dirac modes, 
which are essential for chiral symmetry breaking.
From both analytical and numerical results,
we conclude that there is no one-to-one 
correspondence between confinement and chiral symmetry breaking.
\begin{table}[htb]
\caption{
Numerical results for $\langle L_P \rangle$ and 
$\langle L_P \rangle_{\rm IR\hbox{-}cut}$ 
in lattice QCD with $10^3\times5$ and $\beta=5.6$ 
for each gauge configuration, 
where the system is in confinement phase.
}
  \begin{tabular}{|c|c|c|c|c|c|c|c|} \hline
configuration No. &1&2&3&4&5&6&7 \\ \hline
Re$\langle L_P \rangle$             &0.00961 &-0.00161&0.0139    &-0.00324&0.000689&0.00423&-0.00807\\ 
Im$\langle L_P \rangle$             &-0.00322&-0.00125&-0.00438&-0.00519&-0.0101  &-0.0168&-0.00265 \\\hline 
Re$\langle L_P \rangle_{\rm IR\hbox{-}cut}$ &0.00961 &-0.00160&0.0139    &-0.00325&0.000706&0.00422&-0.00807 \\ 
Im$\langle L_P \rangle_{\rm IR\hbox{-}cut}$ &-0.00321&-0.00125&-0.00437&-0.00520&-0.0101  &-0.0168&-0.00264 \\\hline
  \end{tabular}
\end{table}
\begin{table}[htb]
\caption{
Numerical results for $\langle L_P \rangle$ and 
$\langle L_P \rangle_{\rm IR\hbox{-}cut}$ 
in lattice QCD with $10^3\times3$ and $\beta=5.7$ for each gauge configuration, 
where the system is in deconfinement phase.
}
  \begin{tabular}{|c|c|c|c|c|c|c|c|} \hline
configuration No. &1&2&3&4&5&6&7 \\ \hline
Re$\langle L_P \rangle$             &0.316    &0.337     &0.331    &0.305     &0.314   &0.316      &0.337 \\
Im$\langle L_P \rangle$              &-0.00104&-0.00597&0.00723 &-0.00334&0.00167&0.000120&0.0000482 \\\hline
Re$\langle L_P \rangle_{\rm IR\hbox{-}cut}$ &0.319    &0.340     &0.334    &0.307     &0.317   &0.319      &0.340\\
Im$\langle L_P \rangle_{\rm IR\hbox{-}cut}$ &-0.00103&-0.00597&0.00724 &-0.00333&0.00167&0.000121 &0.0000475\\\hline
  \end{tabular}
\end{table}
\section{Summary and concluding remarks}
In this study, we have analytically derived a direct relation connecting 
the Polyakov loop and the Dirac modes in temporally odd-number lattice QCD, 
with the normal (nontwisted) periodic boundary condition.
We have shown that the low-lying Dirac modes have little 
contribution to the Polyakov loop, which means no one-to-one correspondence 
between confinement and chiral symmetry breaking in QCD.
As a new method, we have modified the KS formalism to 
perform the spin-diagonalizing of the Dirac operator 
on the temporally odd-number lattice.
%matrix $M$ defined by Eq.(\ref{Eq:M}) as in Eq.(\ref{Eq:MDiracM}).
Using the modified KS formalism, we have numerically shown that 
the contribution of low-lying Dirac modes to the Polyakov loop 
is negligible in both confinement and deconfinement phases.

%As future work, we investigate the property of factor 
%$\langle n|\hat{U}_4| n \rangle$ and the behavior of Eq.(\ref{Eq:RelOrig}) 
%at continuum limit and thermodynamic limit.

\vspace{-0.13cm}

\section*{Acknowledgements}

\vspace{-0.13cm}

H.S. and T.I. are supported in part by the Grant for Scientific Research 
[(C) No.23540306, 
%Priority Areas ``New Hadrons'' 
E01:21105006, No.21674002] from the Ministry of Education, 
Science and Technology of Japan.  
The lattice QCD calculation has been done on NEC-SX8R at Osaka University.

\end{document}